\documentclass[twocolumn,A4]{article}
\usepackage[dvips]{graphics}
\usepackage{setspace}
\usepackage{color}
\definecolor{gold}{rgb}{0.85,0.66,0}
\definecolor{dblue}{rgb}{0,0,0.8}
\begin{document}
\onecolumn
\begin{center}
{\bf{\Large {\textcolor{gold}{NAND gate response in a mesoscopic ring: 
An exact study}}}}\\
~\\
{\textcolor{dblue}{Santanu K. Maiti}}$^{1,2,*}$ \\
~\\
{\em $^1$Theoretical Condensed Matter Physics Division,
Saha Institute of Nuclear Physics, \\
1/AF, Bidhannagar, Kolkata-700 064, India \\
$^2$Department of Physics, Narasinha Dutt College,
129, Belilious Road, Howrah-711 101, India} \\
~\\
{\bf Abstract}
\end{center}
NAND gate response in a mesoscopic ring threaded with a magnetic flux 
$\phi$ is investigated by using Green's function formalism. The ring 
is attached symmetrically to two semi-infinite one-dimensional metallic 
electrodes and two gate voltages, namely, $V_a$ and $V_b$, are applied in 
one arm of the ring those are treated as the two inputs of the NAND gate. 
We use a simple tight-binding model to describe the system and numerically 
compute the conductance-energy and current-voltage characteristics as 
functions of the gate voltages, ring-to-electrode coupling strength 
and magnetic flux. Our theoretical study shows that, for $\phi=\phi_0/2$ 
($\phi_0=ch/e$, the elementary flux-quantum) a high output current ($1$) 
(in the logical sense) appears if one or both the inputs to the gate are 
low ($0$), while if both the inputs to the gate are high ($1$), a low 
output current ($0$) appears. It clearly exhibits the NAND gate 
behavior and this feature may be utilized in designing an electronic 
logic gate. 

\vskip 1cm
\begin{flushleft}
{\bf PACS No.}: 73.23.-b; 73.63.Rt. \\
~\\
{\bf Keywords}: Mesoscopic ring; Conductance; $I$-$V$ characteristic;
NAND gate.
\end{flushleft}
\vskip 4.3in
\noindent
{\bf ~$^*$Corresponding Author}: Santanu K. Maiti

Electronic mail: santanu.maiti@saha.ac.in
\newpage
\twocolumn

\section{Introduction}

Electronic transport in quantum confined geometries has attracted
much attention since these simple looking systems are the promising 
building blocks for designing nanodevices especially in electronic 
as well as spintronic engineering. A mesoscopic metallic ring is one
such promising example where electronic motion is restricted. The 
recent progress in nanoscience and technology has allowed us to 
investigate the electron transport through a mesoscopic ring in a 
very tunable way. Using a mesoscopic ring we can make a device that 
can act as a logic gate, which may be used in nanoelectronic circuits. 
To explore this phenomenon we design a bridge system where the ring 
is attached to two external electrodes, so-called the 
electrode-ring-electrode bridge. The ring is then subjected to a 
magnetic flux $\phi$, so-called the Aharonov-Bohm (AB) flux which 
is the key controlling factor for the whole logical operation in 
this particular geometry. The theoretical description of electron 
transport in a bridge system has got much progress following the 
pioneering work of Aviram and Ratner~\cite{aviram}. 
Later, many excellent experiments~\cite{reed1,reed2,tali} have been 
done in several bridge systems to understand the basic mechanisms 
underlying the electron transport. Though in literature both 
theoretical~\cite{nitzan1,nitzan2,orella1,orella2,muj1,muj2,walc2,
walc3,cui} as well as experimental~\cite{reed1,reed2,tali} works on 
electron transport are available, yet lot of controversies are still 
present between the theory and experiment, and the complete knowledge 
of the conduction mechanism in this scale is not very well established 
even today. The interface geometry between the ring and the electrodes
significantly controls the electronic transport in the ring. By changing 
the geometry, one can tune the transmission probability of an electron 
across the ring which is solely due to the effect of quantum interference 
among the electronic waves passing through different arms of the ring. 
Furthermore, the electron transport in the ring can be modulated in 
other way by tuning the magnetic flux, that threads the ring. The AB 
flux threading the ring may change 
the phases of the wave functions propagating along the different arms 
of the ring leading to constructive or destructive interferences, and 
therefore, the transmission amplitude changes~\cite{baer2,baer3,tagami,
walc1,baer1}. Beside these factors, ring-to-electrode coupling is 
another important issue that controls the electron transport in a 
meaningful way~\cite{baer1}. All these are the key factors which 
regulate the electron transmission in the electrode-ring-electrode 
bridge system and these effects have to be taken into account properly 
to reveal the transport mechanisms. 

The main focus of the present work is to describe the NAND gate 
response in a mesoscopic ring threaded by a magnetic flux $\phi$. 
The ring is contacted symmetrically to the electrodes, and the two 
gate voltages $V_a$ and $V_b$ are applied in one arm of the ring 
(see Fig.~\ref{nand}) those are treated as the two inputs of the 
NAND gate. A simple tight-binding model is used to describe the 
system and all the calculations are done numerically. Here we 
address the NAND gate behavior by studying the conductance-energy and 
current-voltage characteristics as functions of the ring-electrodes 
coupling strengths, magnetic flux and gate voltages. Our study reveals 
that for a particular value of the magnetic flux, $\phi=\phi_0/2$, a high 
output current ($1$) (in the logical sense) is available if one or both
the inputs to the gate are low ($0$), while if both the inputs to the
gate are high ($1$), a low output current ($0$) appears. This phenomenon 
clearly shows the NAND gate behavior. To the best of our knowledge the 
NAND gate response in such a simple system has yet not been addressed 
in the literature.

The scheme the paper is as follow. Following the introduction 
(Section $1$), in Section $2$, we described the model and the 
theoretical formulations for the calculation. Section $3$ explores 
the results, and finally, we conclude our study in Section $4$.

\section{Model and the synopsis of the theoretical background}

\begin{figure}[ht]
{\centering \resizebox*{7cm}{4.7cm}{\includegraphics{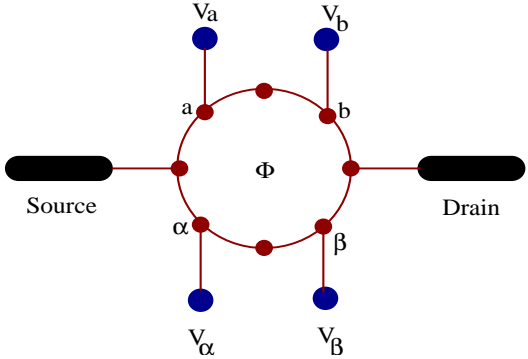}}\par}
\caption{(Color online). Schematic representation for the operation of 
a NAND gate. The atoms $a$, $b$, $\alpha$ and $\beta$ are subjected to 
the voltages $V_a$, $V_b$, $V_{\alpha}$ and $V_{\beta}$, respectively, 
those are variable.}
\label{nand}
\end{figure}

Let us begin by referring to Fig.~\ref{nand}. A mesoscopic ring, 
threaded by a magnetic flux $\phi$, is attached symmetrically 
(upper and lower arms have equal number of lattice points) to 
two semi-infinite one-dimensional ($1$D) metallic electrodes. 
The atoms $a$ and $b$ in the upper arm of the ring are subjected 
to the gate voltages $V_a$ and $V_b$, respectively, those are 
treated as the two inputs of the NAND gate. On the other hand,
two additional voltages $V_{\alpha}$ and $V_{\beta}$ are applied
in the atoms $\alpha$ and $\beta$, respectively, in the lower arm
of the ring.

At very low temperature and bias voltage the conductance $g$ of 
the ring can be expressed from the Landauer conductance 
formula~\cite{datta,marc},
\begin{equation}
g=\frac{2e^2}{h} T
\label{equ1}
\end{equation}
where $T$ gives the transmission probability of an electron across 
the ring. This $(T)$ can be represented in terms of the Green's 
function of the ring and its coupling to the two electrodes by the 
relation~\cite{datta,marc},
\begin{equation}
T={\mbox{Tr}}\left[\Gamma_S G_{R}^r \Gamma_D G_{R}^a\right]
\label{equ2}
\end{equation}
where $G_{R}^r$ and $G_{R}^a$ are respectively the retarded and 
advanced Green's functions of the ring including the effects of 
the electrodes. Here $\Gamma_S$ and $\Gamma_D$ describe the 
coupling of the ring to the source and drain, respectively. For 
the complete system i.e., the ring, source and drain, the Green's 
function is defined as,
\begin{equation}
G=\left(E-H\right)^{-1}
\label{equ3}
\end{equation}
where $E$ is the injecting energy of the source electron. To Evaluate
this Green's function, the inversion of an infinite matrix is needed since
the complete system consists of the finite ring and the two semi-infinite 
electrodes. However, the entire system can be partitioned into sub-matrices 
corresponding to the individual sub-systems and the Green's function for 
the ring can be effectively written as,
\begin{equation}
G_{R}=\left(E-H_{R}-\Sigma_S-\Sigma_D\right)^{-1}
\label{equ4}
\end{equation}
where $H_{R}$ is the Hamiltonian of the ring that can be expressed within 
the non-interacting picture like,
\begin{eqnarray}
H_{R} & = & \sum_i \left(\epsilon_{i0} + V_a \delta_{ia} + V_b \delta_{ib} 
+ V_{\alpha} \delta_{i\alpha} \right. \nonumber \\
 & & \left. + V_{\beta} \delta_{i\beta} \right) + \sum_{<ij>} t 
\left(c_i^{\dagger} c_j e^{i\theta}+ h.c. \right)
\label{equ5}
\end{eqnarray}
In this Hamiltonian $\epsilon_{i0}$'s are the site energies for all the 
sites $i$ except the sites $i=a$, $b$, $\alpha$ and $\beta$ where the 
gate voltages $V_a$, $V_b$, $V_{\alpha}$ and $V_{\beta}$ are applied, 
those are variable. These gate voltages can be incorporated through 
the site energies as expressed in the above Hamiltonian. $c_i^{\dagger}$ 
($c_i$) is the creation (annihilation) operator of an electron at the 
site $i$ and $t$ is the nearest-neighbor hopping integral. The phase 
factor $\theta=2 \pi \phi/N \phi_0$ comes due to the flux $\phi$ 
threaded by the ring, where $N$ corresponds to the total number of 
atomic sites in the ring. In Eq.~(\ref{equ4}),
$\Sigma_S=h_{SR}^{\dagger}g_S h_{SR}$ and
$\Sigma_D=h_{DR} g_D h_{DR}^{\dagger}$ are the self-energy operators due 
to the two electrodes, where $g_S$ and $g_D$ are the Green's functions 
for the source and drain, respectively. $h_{SR}$ and $h_{DR}$ are the 
coupling matrices and they will be non-zero only for the adjacent points 
of the ring and the electrodes as shown in Fig.~\ref{nand}. The coupling 
terms $\Gamma_S$ and $\Gamma_D$ for the ring can be calculated through 
the expression~\cite{datta,tian},
\begin{equation}
\Gamma_{S(D)}=i\left[\Sigma_{S(D)}^r-\Sigma_{S(D)}^a\right]
\end{equation}
where $\Sigma_{S(D)}^r$ and $\Sigma_{S(D)}^a$ are the retarded and
advanced self-energies, respectively, and they are conjugate to each
other. Datta {\em et al.}~\cite{datta,tian} have shown that the 
self-energies can be expressed like as,
\begin{equation}
\Sigma_{S(D)}^r=\Lambda_{S(D)}-i \Delta_{S(D)}
\end{equation}
where $\Lambda_{S(D)}$ are the real parts of the self-energies which
correspond to the shift of the energy eigenvalues of the ring, and
the imaginary parts $\Delta_{S(D)}$ of the self-energies represent the
broadening of these energy levels. This broadening is much larger than 
the thermal broadening, and accordingly, we restrict our all calculations 
only at absolute zero temperature. These two self-energies $\Sigma_S$ 
and $\Sigma_D$ bear all the information of the coupling of the ring 
to the source and drain, respectively~\cite{datta}. A similar kind of 
tight-binding Hamiltonian is also used, except the phase factor 
$\theta$, to describe the $1$D perfect electrodes where the 
Hamiltonian is parametrized by constant on-site potential $\epsilon_0$ 
and nearest-neighbor hopping integral $t_0$. The hopping integral between
the source and the ring is $\tau_S$, while it is $\tau_D$ between the
ring and the drain. 

The current $I$ passing through the ring is depicted as a single-electron
scattering process between the two reservoirs of charge carriers. The
current-voltage relation is evaluated from the following
expression~\cite{datta},
\begin{equation}
I(V)=\frac{e}{\pi \hbar}\int \limits_{E_F-eV/2}^{E_F+eV/2} T(E,V) dE
\label{equ8}
\end{equation}
where $E_F$ is the equilibrium Fermi energy. Here we assume that the 
entire voltage is dropped across the ring-electrode interfaces, and it 
is examined that under such an assumption the $I$-$V$ characteristics 
do not change their qualitative features. 

All the results in this communication are determined at absolute
zero temperature, but they should valid even for finite temperature
($\sim 300$ K), since the broadening of the energy levels of the ring 
due to its coupling to the electrodes becomes much larger than that of
the thermal broadening~\cite{datta}. For simplicity, we take the unit 
$c=e=h=1$ in our present calculation. 

\section{Results and discussion}

Let us start our discussion by mentioning the values of the 
different parameters used for the numerical calculation. In the ring, 
\begin{figure}[ht]
{\centering \resizebox*{8cm}{7cm}{\includegraphics{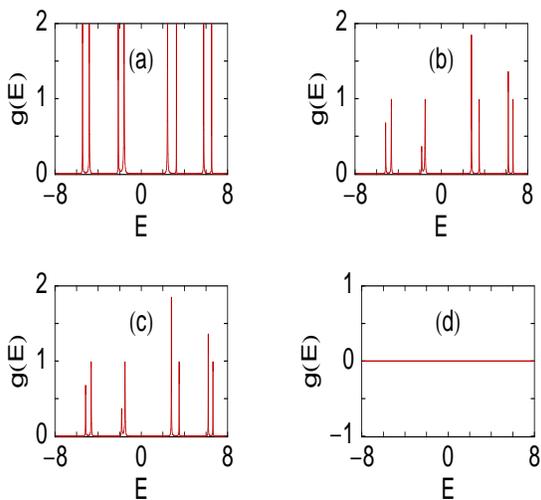}}\par}
\caption{(Color online). $g$-$E$ curves in the weak-coupling limit for a 
mesoscopic ring with $N=8$, $V_{\alpha}=V_{\beta}=2$ and $\phi=0.5$.
(a) $V_a=V_b=0$, (b) $V_a=2$ and $V_b=0$, (c) $V_a=0$ and $V_b=2$ and
(d) $V_a=V_b=2$.}
\label{condlow}
\end{figure}
the on-site energy $\epsilon_{i0}$ is taken as $0$ for all the sites 
$i$, except the sites $i=a$, $b$, $\alpha$ and $\beta$ where the site 
energies are taken as $V_a$, $V_b$, $V_{\alpha}$ and $V_{\beta}$ 
respectively, and the nearest-neighbor hopping strength $t$ is set
to $3$. On the other hand, for the side attached electrodes the 
on-site energy ($\epsilon_0$) and the nearest-neighbor hopping strength 
($t_0$) are fixed to $0$ and $4$, respectively. The voltages $V_{\alpha}$
and $V_{\beta}$ are set to $2$ and the Fermi energy $E_F$ is set at $0$. 
Throughout the study, we 
focus our results for the two limiting cases depending on the strength 
of the coupling of the ring to the source and drain. In one case we use 
the condition $\tau_{S(D)} << t$, which is so-called the weak-coupling 
limit. For this regime we choose $\tau_S=\tau_D=0.5$. In the other case
the condition $\tau_{S(D)} \sim t$ is used, which is named as the 
strong-coupling limit. In this particular regime, the values of the 
parameters are set as $\tau_S=\tau_D=2.5$. The key controlling 
parameter for all these calculations is the magnetic flux $\phi$ 
which is set to $\phi_0/2$ i.e., 0.5 in our chosen unit.

Figure~\ref{condlow} shows the variation of the conductance ($g$) as
a function of the injecting electron energy ($E$) in the limit of
\begin{figure}[ht]
{\centering \resizebox*{8cm}{7cm}{\includegraphics{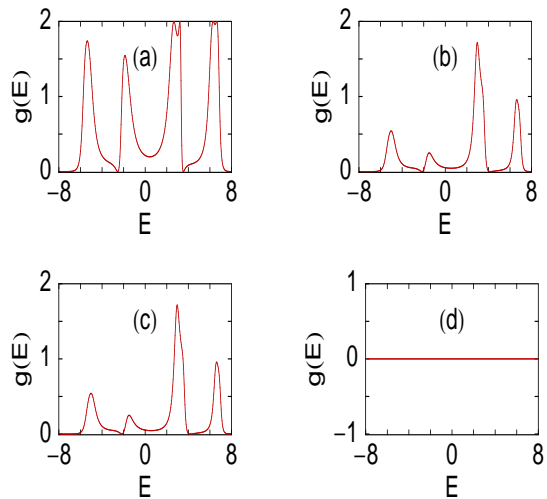}}\par}
\caption{(Color online). $g$-$E$ curves in the strong-coupling limit for a 
mesoscopic ring with $N=8$, $V_{\alpha}=V_{\beta}=2$ and $\phi=0.5$.
(a) $V_a=V_b=0$, (b) $V_a=2$ and $V_b=0$, (c) $V_a=0$ and $V_b=2$ and
(d) $V_a=V_b=2$.}
\label{condhigh}
\end{figure}
weak-coupling for a mesoscopic ring with $N=8$ and $V_{\alpha}=V_{\beta}=2$,
where (a), (b), (c) and (d) correspond to the results for the different
values of $V_a$ and $V_b$. When both the two inputs $V_a$ and $V_b$
are identical to $2$ i.e., both the inputs are high, the conductance
$g$ becomes exactly zero (Fig.~\ref{condlow}(d)) for all energies.
This reveals that the electron cannot conduct from the source to the
drain through the ring. While, for the cases where anyone or both the 
inputs to the gate are zero (low), the conductance shows fine resonant 
peaks for some particular energies, as shown in Figs.~\ref{condlow}(a), 
(b) and (c), respectively. Thus, in all these three cases, the electron 
can conduct through the ring. From Fig.~\ref{condlow}(a) it is observed 
that, at the resonances the conductance $g$ approaches the value $2$, 
and accordingly, the transmission probability $T$ becomes unity, since 
the expression $g=2T$ is satisfied from the Landauer conductance formula 
(see Eq.~\ref{equ1} with $e=h=1$). The height of the resonant peaks
gets down ($T<1$) for the cases where anyone of the two inputs is high 
and other is low (Figs.~\ref{condlow}(b) and (c)). All these resonant
peaks are associated with the energy eigenvalues of the ring, and thus, 
it can be predicted that the conductance spectrum manifests itself the 
\begin{figure}[ht]
{\centering \resizebox*{8cm}{7cm}{\includegraphics{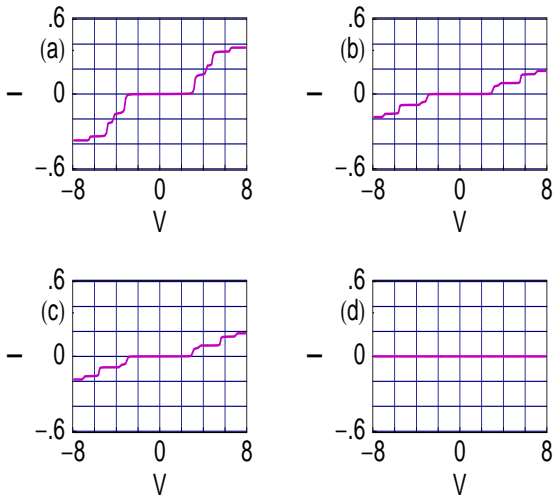}}\par}
\caption{(Color online). $I$-$V$ curves in the weak-coupling limit for a 
mesoscopic ring with $N=8$, $V_{\alpha}=V_{\beta}=2$ and $\phi=0.5$.
(a) $V_a=V_b=0$, (b) $V_a=2$ and $V_b=0$, (c) $V_a=0$ and $V_b=2$ 
and (d) $V_a=V_b=2$.}
\label{currlow}
\end{figure}
electronic structure of the ring. This reveals that more resonant peaks
are expected for the larger rings, associated with their energy spectra.
Now we illustrate the dependences of the gate voltages on the electron
transport for these four different cases. The probability amplitude of
getting an electron across the ring depends on the quantum interference of 
the electronic waves passing through the two arms (upper and lower) of the
ring. For the symmetrically connected ring i.e., when the two arms of the
ring become identical with each other, the probability amplitude is exactly
zero ($T=0$) for the flux $\phi=\phi_0/2$. This is due to the result of
the quantum interference among the two waves in the two arms of the ring,
which can be obtained in a very simple mathematical calculation. So, for
the case when both the two inputs to the gate are identical to $2$ i.e., 
$V_a=V_b=2$, the upper and lower arms of the ring become exactly similar. 
This is due to the fact that the potentials $V_{\alpha}$ and $V_{\beta}$ 
are also fixed to $2$. Therefore, in this particular case the transmission 
probability drops to zero. If the two inputs $V_a$ and $V_b$ are different 
from the potentials applied in the sites $\alpha$ and $\beta$, then the 
two arms are no longer identical to each other and the transmission
probability will not vanish. Hence, to get the zero transmission 
probability when both the inputs are high, we should tune $V_{\alpha}$ 
and $V_{\beta}$ properly, observing the input potentials and vice versa.
On the other hand, due to the breaking of the symmetry of the two arms
(i.e., two arms are no longer identical to each other) in the other 
three cases by making anyone or both the two inputs to the gate are 
zero (low), the non-zero value of the transmission probability is 
achieved which
\begin{table}[ht]
\begin{center}
\caption{NAND gate behavior in the limit of weak-coupling. The current 
$I$ is computed at the bias voltage $6.02$.}
\label{table1}
~\\
\begin{tabular}{|c|c|c|}
\hline \hline
Input-I ($V_a$) & Input-II ($V_b$) & Current ($I$) \\ \hline 
$0$ & $0$ & $0.339$ \\ \hline
$2$ & $0$ & $0.157$ \\ \hline
$0$ & $2$ & $0.157$ \\ \hline
$2$ & $2$ & $0$ \\ \hline \hline
\end{tabular}
\end{center}
\end{table}
reveals the electron conduction across the ring. The reduction of the 
transmission probability from unity for the cases where any one of the 
two gates is high and other is low compared to the case where both the 
gates are low is solely due to the quantum interference effect. Thus 
it can be emphasized that the electron conduction takes place across 
the ring if any one or both the inputs to the gate are low, while if 
both the inputs are high, the conduction is no longer possible. This 
feature clearly describes the NAND gate behavior. In this context we 
also discuss the effect of the ring-to-electrode coupling. As 
illustrative examples, in Fig.~\ref{condhigh} we show the 
conductance-energy characteristics for the strong-coupling limit, 
where (a), (b), (c) and (d) are drawn respectively for the same gate 
voltages as in Fig.~\ref{condlow}. In the strong-coupling case, all the 
resonant peaks get substantial widths compared to the weak-coupling case. 
This is due to the broadening of the energy levels of the ring in the
limit of strong coupling, where the contribution comes from the 
imaginary parts of the self-energies $\Sigma_S$ and $\Sigma_D$, 
respectively~\cite{datta}. Therefore, by tuning the coupling strength, 
we can get the electron transmission across the ring for the wider 
range of energies and it provides an important behavior in the study 
of current-voltage ($I$-$V$) characteristics.

All these properties of electron transfer can be much more clearly 
understood from our presented current-voltage ($I$-$V$) characteristics.
The current across the ring is determined by integrating the
transmission function $T$ as prescribed in Eq.~\ref{equ8}. The 
transmission function varies exactly similar to that of the conductance 
spectrum, differ only in magnitude by the factor $2$ since the relation 
$g=2T$ holds from the Landauer conductance formula Eq.~\ref{equ1}.
As representative examples, in Fig.~\ref{currlow} we plot the 
$I$-$V$ characteristics for a mesoscopic ring with $N=8$ in the limit 
\begin{figure}[ht]
{\centering \resizebox*{8cm}{7cm}{\includegraphics{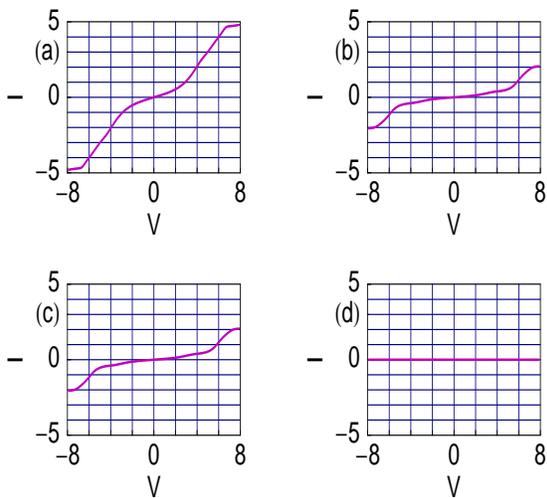}}\par}
\caption{(Color online). $I$-$V$ curves in the strong-coupling limit for 
a mesoscopic ring with $N=8$, $V_{\alpha}=V_{\beta}=2$ and $\phi=0.5$.
(a) $V_a=V_b=0$, (b) $V_a=2$ and $V_b=0$, (c) $V_a=0$ and $V_b=2$ 
and (d) $V_a=V_b=2$.}
\label{currhigh}
\end{figure}
of weak-coupling, where (a), (b), (c) and (d) correspond to the results
for the different cases of the input voltages. Quite interestingly we
see that when both the two inputs to the gate are identical to $2$
(high), the current $I$ becomes zero (see Fig.~\ref{currlow}(d)) for the
entire bias voltage $V$. This feature is clearly understood from the
conductance spectrum, Fig.~\ref{condlow}(d), since the current is
computed from the integration method of the transmission function
$T$. In all the other three cases of the input voltages, a non-zero 
value of the current is obtained, those are clearly presented in 
Figs.~\ref{currlow}(a), (b) and (c). These figures show that
the current exhibits staircase-like structure with fine steps as a
function of the applied bias voltage. This is due to the existence of
the fine resonant peaks in the conductance spectra in the weak-coupling
limit. With the increase of the bias voltage $V$,
the electrochemical potentials on the electrodes are shifted gradually,
and finally cross one of the quantized energy levels of the ring.
Accordingly, a current channel is opened up which provides a jump
in the $I$-$V$ characteristic curve. Here, it is also important 
to mention that the non-zero value of the current appears beyond a 
finite value of the bias voltage $V$, so-called the threshold 
voltage ($V_{th}$). This $V_{th}$ can be changed by controlling 
the size ($N$) of the ring. From these current-voltage curves, 
the NAND gate behavior of the ring can be observed very nicely.
To make it much clear, in Table~\ref{table1}, we show a quantitative 
estimate of the typical current amplitude determined at the bias 
voltage $V=6.02$. It is observed that, when any one of the two gates 
is high and other is low, the current gets the value $0.157$, and 
\begin{table}[ht]
\begin{center}
\caption{NAND gate behavior in the limit of strong-coupling. The current 
$I$ is computed at the bias voltage $6.02$.}
\label{table2}
~\\
\begin{tabular}{|c|c|c|}
\hline \hline
Input-I ($V_a$) & Input-II ($V_b$) & Current ($I$) \\ \hline 
$0$ & $0$ & $4.018$ \\ \hline
$2$ & $0$ & $1.174$ \\ \hline
$0$ & $2$ & $1.174$ \\ \hline
$2$ & $2$ & $0$ \\ \hline \hline
\end{tabular}
\end{center}
\end{table}
for the case when both the two inputs are low, it ($I$) achieves 
the value $0.339$. While, for the case when both the two inputs are 
high ($V_a=V_b=2$), the current becomes exactly zero. In the same 
footing, as above, here we also describe the $I$-$V$ characteristics 
for the strong-coupling limit. In this limit, the current varies 
almost continuously with the applied bias voltage and gets much larger 
amplitude than the weak-coupling case as presented in Fig.~\ref{currhigh}. 
The reason is that, in the limit of strong-coupling all the energy 
levels get broadened which provide larger current in the integration 
procedure of the transmission function $T$. Thus by tuning the
strength of the ring-to-electrode coupling, we can achieve very large
current amplitude from the very low one for the same bias voltage $V$.
All the other properties i.e., the dependences of the gate voltages on the
$I$-$V$ characteristics are exactly similar to those as given in
Fig.~\ref{currlow}. In this strong-coupling limit we also make a 
quantitative study for the typical current amplitude, given in 
Table~\ref{table2}, where the current amplitude is determined at 
the same bias voltage ($V=6.02$) as earlier. The response of the 
output current is exactly similar to that as given in Table~\ref{table1}. 
Here the current achieves the value $1.174$ in the cases where any one 
of the two gates is high and other is low, and it ($I$) becomes $4.018$ 
for the case where both the two inputs are low. On the other hand, the 
current becomes exactly zero for the case where $V_a=V_b=2$. The 
non-zero values of the current in this strong-coupling limit are much 
larger than the weak-coupling case, as expected. From these results 
we can clearly manifest that a mesoscopic ring exhibits the NAND gate 
response.

\section{Concluding remarks}

In this work, we have explored the NAND gate response in a mesoscopic 
metallic ring threaded by a magnetic flux $\phi$ in the Green's
function formalism. The ring is attached symmetrically to the electrodes
and two gate voltages $V_a$ and $V_b$ are applied in one arm of the
ring those are taken as the two inputs of the NAND gate. A simple
tight-binding model is used to describe the system and all the 
calculations are exact and performed numerically. We have computed the 
conductance-energy and current-voltage characteristics as functions 
of the gate voltages, ring-to-electrode coupling strength and 
magnetic flux. Very interestingly we have observed that, for the half 
flux-quantum value of $\phi$ ($\phi=\phi_0/2$), a high output current 
($1$) (in the logical sense) appears if one or both the inputs to 
the gate are low ($0$). On the other hand, if both the inputs to the
gate are high ($1$), a low output current ($0$) appears. It 
clearly manifests the NAND gate behavior and this aspect may be 
utilized in designing a tailor made electronic logic gate. 

Throughout our work, we have addressed the conductance-energy and
current-voltage characteristics for a quantum ring with total
number of atomic sites $N=8$. In our model calculations, this typical
number ($N=8$) is chosen only for the sake of simplicity. Here it
is important to note that for a larger ring with more atomic sites,
one can also apply input voltages to the other atomic sites, preserving
the symmetry between the upper and lower arms of the ring. Though the
results presented here change numerically with the ring size ($N$)
and the distance between the input positions, but all the basic 
features remain exactly invariant since the results solely depend on
the quantum interference effect of the electronic waves passing through 
the two arms of the ring. To be more specific, it is important to note 
that, in real situation the experimentally achievable rings have typical 
diameters within the range $0.4$-$0.6$ $\mu$m. In such a small ring, 
unrealistically very high magnetic fields are required to produce a 
quantum flux. To overcome this situation, Hod {\em et al.} have studied 
extensively and proposed how to construct nanometer scale devices, based 
on Aharonov-Bohm interferometry, those can be operated in moderate magnetic 
fields~\cite{baer4,baer5,baer6,baer7}.

In the present paper we have done all the calculations by ignoring
the effects of the temperature, electron-electron correlation, disorder,
etc. Due to these factors, any scattering process that appears in the
arms of the rings would have influence on electronic phases, and, in
consequences can disturb the quantum interference effects. Here we
have assumed that, in our sample all these effects are too small, and
accordingly, we have neglected all these factors in this particular
study.

The importance of this article is concerned with (i) the simplicity of the
geometry and (ii) the smallness of the size. To the best of our knowledge 
the NAND gate response in such a simple low-dimensional system that can be 
operated even at finite temperature ($\sim 300$ K) has not been 
addressed earlier in the literature.

\end{document}